\documentclass[iop,apjl]{emulateapj}

\newcommand{\q}[1]{_{\rm #1}}
\newcommand{\ten}[1]{$10^{#1}$}
\newcommand{\scit}[2]{$#1\times10^{#2}$}
\newcommand{\scim}[2]{#1\times10^{#2}}

\newcommand{\ps}{s$^{-1}$}
\newcommand{\pcs}{cm$^{-2}$}
\newcommand{\pcc}{cm$^{-3}$}
\newcommand{\kkms}{K km s$^{-1}$}
\newcommand{\mkkms}{mK km s$^{-1}$}
\newcommand{\eq}[1]{Equation \ref{eq:#1}}
\newcommand{\fig}[1]{Figure \ref{fig:#1}}
\newcommand{\tb}[1]{Table \ref{tb:#1}}
\newcommand{\sect}[1]{Section\ \ref{sec:#1}}
\newcommand{\mh}{H$_2$}
\newcommand{\w}{H$_2$O}
\newcommand{\ws}{H$_2^{16}$O}
\newcommand{\wsev}{H$_2^{17}$O}
\newcommand{\we}{H$_2^{18}$O}
\newcommand{\ceo}{C$^{18}$O}
\newcommand{\fmh}{H$_2$CO}
\newcommand{\meoh}{CH$_3$OH}
\newcommand{\mecn}{CH$_3$CN}
\newcommand{\wline}{$3_{12}$--$3_{03}$}
\newcommand{\pdbi}{$3_{13}$--$2_{20}$}
\newcommand{\eup}{E_{\rm u}}

\newcommand{\tmbdv}{\int T_{\rm mb}{\rm d}v}

\newcommand{\cdws}{N({\rm H}_2^{16}{\rm O})}

\newcommand{\cdwe}{N({\rm H}_2^{18}{\rm O})}
\newcommand{\cdceo}{N({\rm C}^{18}{\rm O})}

\slugcomment{Accepted by ApJ -- March 25, 2013}

\shorttitle{Hot water in NGC1333 IRAS2A}
\shortauthors{Visser et al.}

\begin{document}

\title{Hot water in the inner 100 AU of the Class 0 protostar NGC1333 IRAS2A}

\author{Ruud Visser\altaffilmark{1}, Jes K. J{\o}rgensen\altaffilmark{2,3}, Lars E. Kristensen\altaffilmark{4,5}, Ewine F. van Dishoeck\altaffilmark{4,6}, and Edwin A. Bergin\altaffilmark{1}}
\altaffiltext{1}{Department of Astronomy, University of Michigan, 500 Church Street, Ann Arbor, MI 48109-1042, USA; visserr@umich.edu}
\altaffiltext{2}{Niels Bohr Institute, University of Copenhagen, Juliane Maries Vej 30, 2100 Copenhagen {\O}, Denmark}
\altaffiltext{3}{Centre for Star and Planet Formation, Natural History Museum of Denmark, University of Copenhagen, {\O}ster Voldgade 5-7, 1350 Copenhagen K, Denmark}
\altaffiltext{4}{Leiden Observatory, Leiden University, P.O.\ Box 9513, 2300 RA Leiden, the Netherlands}
\altaffiltext{5}{Harvard-Smithsonian Center for Astrophysics, 60 Garden Street, Cambridge, MA 02138, USA}
\altaffiltext{6}{Max-Planck-Institut f\"ur Extraterrestrische Physik, Giessenbachstrasse 1, 85748 Garching, Germany}

\begin{abstract}
Evaporation of water ice above 100 K in the inner few 100 AU of low-mass embedded protostars (the so-called hot core) should produce quiescent water vapor abundances of $\sim$\ten{-4} relative to \mh. Observational evidence so far points at abundances of only a few \ten{-6}. However, these values are based on spherical models, which are known from interferometric studies to be inaccurate on the relevant spatial scales. Are hot cores really that much drier than expected, or are the low abundances an artifact of the inaccurate physical models? We present deep velocity-resolved Herschel-HIFI spectra of the \wline{} lines of \ws{} and \we{} (1097 GHz, $\eup/k=249$ K) in the low-mass Class 0 protostar NGC1333 IRAS2A\@. A spherical radiative transfer model with a power-law density profile is unable to reproduce both the HIFI data and existing interferometric data on the \we{} \pdbi{} line (203 GHz, $\eup/k=204$ K). Instead, the HIFI spectra likely show optically thick emission from a hot core with a radius of about 100 AU\@. The mass of the hot core is estimated from the \ceo{} $J=9$--8 and 10--9 lines. We derive a lower limit to the hot water abundance of \scit{2}{-5}, consistent with the theoretical predictions of $\sim$\ten{-4}. The revised HDO/\w{} abundance ratio is \scit{1}{-3}, an order of magnitude lower than previously estimated.
\end{abstract}



\section{Introduction}
\label{sec:intro}
In the embedded stage of low-mass star formation, the central source is surrounded by a collapsing envelope that spans a wide range in densities and temperatures \citep{andre00a}. The inner region, known as the hot core or hot corino \citep[$T>100$ K;][]{walmsley92a,vandishoeck98a,ceccarelli04a}, forms a crucial step in the flow of matter from the cold outer envelope toward the circumstellar disk \citep{visser09a}. Thermal evaporation of water ice at about 100 K should produce a hot core water abundance of about \ten{-4} \citep{ceccarelli96a,rodgers03a}. Such abundances are indeed observed for high-mass protostars \citep{vandertak06a,chavarria10a,herpin12a}, though not universally \citep{emprechtinger13a}. Abundances reported for low-mass hot cores have so far not exceeded \ten{-5} \citep{ceccarelli00a,maret02a,kristensen10b,coutens12a}. Are low-mass hot cores that much ``drier'' than expected?

Measuring the hot core water abundance in a protostar is not a trivial task, regardless of whether it is a low- or a high-mass source. Shocked gas tends to outshine the quiescent inner envelope \citep{melnick00a,chavarria10a,kristensen10b,kristensen12a,coutens12a,emprechtinger13a}, so one has to isolate the envelope emission from velocity-resolved spectra. Hot core abundances derived from spectrally unresolved data, such as from the Infrared Space Observatory \citep[ISO;][]{ceccarelli00a,maret02a}, should therefore be treated with caution.

Another difficulty is the poorly known source structure on the spatial scales of the hot core: even if the column of quiescent hot water can be measured, there is no reliable column of hot \mh{} to compare against. Abundances of water and other molecules are typically computed by way of spherical envelope models with simple density profiles, constrained from single-dish dust continuum observations \citep{ceccarelli00a,vandertak00c,schoier02a,maret02a,coutens12a}. Based on such a model, we previously derived a hot core abundance of $\leq10^{-5}$ for the Class 0 protostar NGC1333 IRAS2A \citep{kristensen10b,liu11a}. However, spherical models do not recover the density enhancements measured with continuum interferometry in the inner few 100 AU of various low-mass protostars \citep{jorgensen04a,jorgensen05b,jorgensen07a,chiang08a}. The low water abundances may thus merely be an artefact of the inaccuracy of the adopted source models on the spatial scales of the hot core.

Using interferometric observations of the \pdbi{} line of \we{} at 203 GHz ($\eup/k=204$ K), \citet{jorgensen10a} and \citet{persson12a} measured hot water column densities for the three Class 0 protostars NGC1333 IRAS2A, 4A, and 4B\@. They also estimated \mh{} column densities from continuum interferometry on the same spatial scales to derive water abundances between a few \ten{-9} and a few \ten{-6}. However, these abundances are relative to the total amount of gas in the inner $\sim$100 AU, not relative to just the gas above 100 K\@. If a significant fraction of the material on these scales resides in an embedded disk or pseudo-disk, the bulk would be at lower temperatures and most water would be frozen out \citep{visser09a,ilee11a}.

In order to get additional constraints on the hot water in NGC1333 IRAS2A, we performed a 5.1-hr integration of the \wline{} lines of \ws, \wsev, and \we{} at 1097 GHz with the Heterodyne Instrument for the Far-Infrared \citep[HIFI;][]{degraauw10a} on the Herschel Space Observatory \citep{pilbratt10a}.\footnote{\emph{Herschel} is an ESA space observatory with science instruments provided by European-led Principal Investigator consortia and with important participation from NASA.} This observation is part of the key program ``Water in star-forming regions with Herschel'' \citep[WISH;][]{vandishoeck11a}, which aims to study the physics and chemistry of water during star formation across a range of masses and evolutionary stages. HIFI has high enough spectral resolution to disentangle the quiescent and shocked gas, and the deep integration allows for a detection of \we\@.

This paper presents the observations (Sections \ref{sec:obs} and \ref{sec:res}) along with a detailed analysis (\sect{disc}). We develop a scenario in which the observed emission originates in the same gas as the 203 GHz line, namely in a hot core with a radius of about 100 AU\@. Our definition of a ``hot core'' is fairly loose and includes all quiescent (i.e., non-shocked) material hotter than 100 K, regardless of whether it is part of the envelope or an embedded disk-like structure. A simple spherical model fails at reproducing both data sets simultaneously, so we perform a model-independent analysis to constrain the distribution of water within the hot core. Comparing the water column density against Herschel-HIFI \ceo{} $J=9$--8 and 10--9 data yields a lower limit on the water abundance of \scit{2}{-5} relative to \mh. As concluded in \sect{conc}, the hot core in IRAS2A does not appear to be as dry as previously estimated from pure spherical models.


\section{Observations and Data Reduction}
\label{sec:obs}
NGC1333 IRAS2A is a low-mass Class 0 protostar \citep[$L\q{bol}=35.7$ $L_\odot$, $T\q{bol}=50$ K;][]{kristensen12a} located at a distance of 235 pc in Perseus at coordinates $3^{\rm h}28^{\rm m}55\fs6$ by $+31\degr14'37\farcs1$ \citep[J2000;][]{hirota08a}. Based on large-scale CO 3--2 maps, it is oriented close to edge on \citep{sandell94a}. IRAS2A was observed on 2011 March 12 (obsid 1342215968) with HIFI on Herschel \citep{pilbratt10a,degraauw10a} in double beam switch mode with a nod of $3'$. The primary targets were the \wline{} lines of ortho-\ws{} and \we{} at 1097.365 and 1095.627 GHz in band 4 ($\eup/k=249$ K). The full spectrum covers a range of 4 GHz in each sideband, including the \wline{} and $1_{11}$--$0_{00}$ lines of \wsev{} and the 10--9 line of \ceo{} (\tb{linelist}). The data were recorded in H and V polarization using both the wide-band spectrometer (WBS; 1.1 MHz resolution) and the high-resolution spectrometer (HRS; 0.5 MHz). The HIFI beam size at 1100 GHz is $19\farcs3$ and the overall flux uncertainty is about 10\% \citep{roelfsema12a}.

The \wline{} line of \ws{} was previously detected in IRAS2A after a 30-min integration \citep{kristensen10b}. The current observation had a total exposure time of 5.1 hr in order to detect the \we{} line and to improve the decomposition of the various velocity components in the \ws{} spectrum (\sect{res}). The longer integration time also puts a more sensitive upper limit on the \wsev{} line.

The data were reduced with HIPE v8.2.1 \citep{ott10a} and exported to CLASS\footnote{\url{http://www.iram.fr/IRAMFR/GILDAS}} for analysis. The intensities were converted from antenna to main-beam temperature scale through a main-beam efficiency of 0.74 \citep{roelfsema12a}. The H and V spectra were averaged after individual inspection. The final data reduction step was to subtract a linear baseline.


\section{Results}
\label{sec:res}


\begin{figure}[t!]
\epsscale{1.17}
\plotone{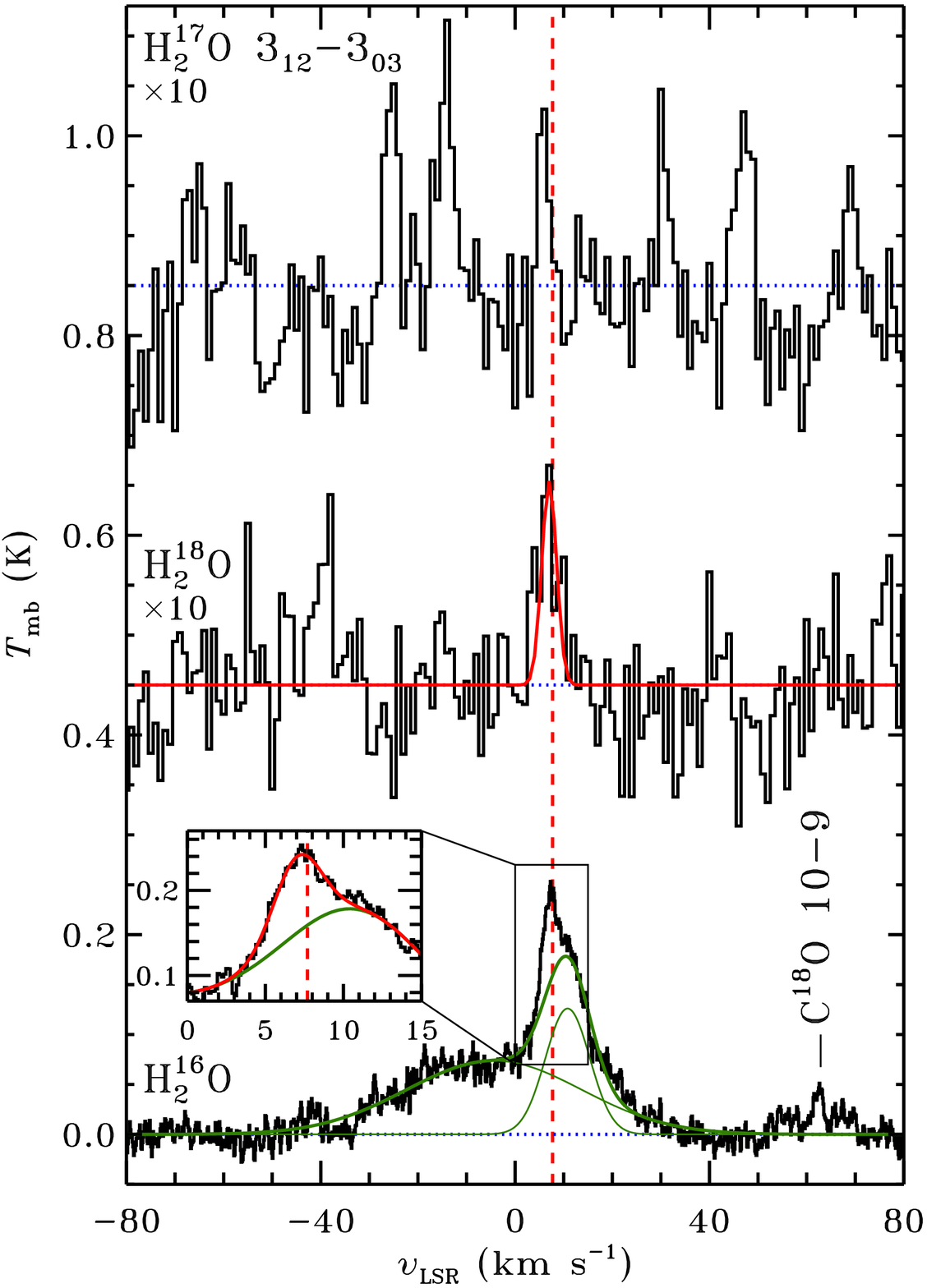}
\caption{Baseline-subtracted spectra of the \wline{} lines of \ws{} (bottom), \we{} (middle), and \wsev{} (top) in NGC1333 IRAS2A\@. The \ws{} spectrum includes Gaussian profiles (green) fitted to the broad and medium components. Shown as an inset is a blow-up of the narrow feature in the \ws{} spectrum, with a Gaussian fitted in red. The narrow \we{} feature is also fitted with a red Gaussian. The vertical dashed red lines mark the source velocity of 7.7 km \ps. The \ws{} spectrum is shown at the original spectral resolution of 0.14 km \ps. The other two spectra are rebinned to 1.0 km \ps{} and blown up by a factor of 10.\label{fig:312spectra}}
\end{figure}

\begin{deluxetable*}{ccccccccccc}
\tablecaption{Integrated Line Intensities ($\tmbdv$ in \kkms) and $3\sigma$ Upper Limits for Narrow Emission in NGC1333 IRAS2A\tablenotemark{a}\label{tb:linelist}}
\tablewidth{0pt}
\tablehead{
\colhead{Species} & \colhead{Transition} & \colhead{Telescope/} & \colhead{$\nu$} & \colhead{$E\q{u}/k$} & \colhead{$A\q{ul}$} & \colhead{$\theta$} & \colhead{Observed}  & \colhead{Model}     & \colhead{Reference\tablenotemark{b}} \\
\colhead{}        & \colhead{}           & \colhead{Instrument} & \colhead{(GHz)} & \colhead{(K)}        & \colhead{\ps} & \colhead{(arcsec)} & \colhead{Intensity} & \colhead{Intensity} & \colhead{}
}
\startdata
p-\ws & $2_{02}$--$1_{11}$ & HIFI &  987.927 & 100.8 & $5.835(-3)$ & 21.5 & $<0.309$ & 0.312 & 1 \\[2pt]
p-\ws & $2_{11}$--$2_{02}$ & HIFI &  752.033 & 136.9 & $7.062(-3)$ & 28.2 & $<0.583$ & 0.230 & 1 \\[2pt]
o-\ws & $3_{12}$--$2_{21}$ & HIFI & 1153.127 & 249.4 & $2.634(-3)$ & 18.4 & $<0.545$ & 0.352 & 1 \\[2pt]
o-\ws & \wline             & HIFI & 1097.365 & 249.4 & $1.648(-2)$ & 19.3 & \phm{$^{\rm b}$}$0.342\pm0.050$\tablenotemark{c} & 0.342 & 2 \\[2pt]
p-\ws & $3_{31}$--$4_{04}$ & HIFI & 1893.687 & 410.4 & $1.630(-4)$ & 11.2 & $<0.531$ & 0.005 & 3 \\[2pt]
p-\we & $2_{02}$--$1_{11}$ & HIFI &  994.675 & 100.6 & $6.020(-3)$ & 21.3 & $<0.068$ & 0.005 & 1 \\[2pt]
p-\we & \pdbi              & PdBI &  203.408 & 203.7 & $4.812(-6)$ & 0.83 & $42\pm8$ & 42 & 4 \\[2pt]
o-\we & \wline             & HIFI & 1095.627 & 248.7 & $1.621(-2)$ & 19.4 & $0.078\pm0.011$ & 0.078 & 2 \\[2pt]
o-\we & $4_{14}$--$3_{21}$ & APEX &  390.608 & 322.0 & $3.143(-5)$ & 16.0 & $<0.410$ & 0.205\tablenotemark{d} & 5 \\[2pt]
o-\we & $4_{23}$--$3_{30}$ & \nodata & 489.054 & 429.6 & $6.887(-5)$ & 0.83 & \nodata & 331  & \nodata \\[2pt]
o-\we & $5_{32}$--$4_{41}$ & \nodata & 692.079 & 727.6 & $1.478(-4)$ & 0.83 & \nodata & 1.74  & \nodata \\[2pt]
p-\wsev & $1_{11}$--$0_{00}$ & HIFI & 1107.167 &  53.1 & $1.812(-2)$ & 19.2 & $<0.037$ & 0.007 & 2 \\[2pt]
o-\wsev & \wline             & HIFI & 1096.414 & 249.1 & $1.635(-2)$ & 19.3 & $<0.043$ & 0.005 & 2 \\[2pt]
\ceo  & 9--8               & HIFI &  987.560 & 237.0 & $7.380(-5)$ & 21.5 & $0.157\pm0.032$ & \phm{$^{\rm c}$}0.043\tablenotemark{e} & 6 \\[2pt]
\ceo  & 10--9\phm{$_0$}    & HIFI & 1097.163 & 289.7 & $9.661(-5)$ & 19.3 & $0.150\pm0.050$ & \phm{$^{\rm c}$}0.035\tablenotemark{e} & 2
\enddata
\tablenotetext{a}{Also listed are the frequency ($\nu$), upper-level energy ($\eup$), Einstein A coefficient ($A\q{ul}$), and beam diameter ($\theta$) for each line. The notation $a(-b)$ denotes \scit{a}{-b}. The model intensities are corrected for beam dilution and dust extinction as described in \sect{dust}.}
\tablenotetext{b}{1: \citet{kristensen10b}; 2: this work; 3: Benz et al.\ (in prep.); 4: \citet{persson12a}; 5: F.\ Wyrowski (priv.\ comm.); 6: \citet{yildiz10a}}
\tablenotetext{c}{$\tmbdv$ for the broad and medium components is $3.34\pm0.09$ and $1.44\pm0.08$ \kkms.}
\tablenotetext{d}{The intensity predicted for a fictitious $0\farcs83$ ALMA beam is 76.4 \kkms}
\tablenotetext{e}{For an \w/CO ratio of unity above 100 K\@. Most of the observed \ceo{} emission originates in gas below 100 K, which is not included in the model.}
\end{deluxetable*}

\subsection{New Data}
\label{sec:newdata}
The full HIFI spectrum as recorded with the WBS backend is shown in Figures \ref{fig:fullspecone} and \ref{fig:fullspec} in the appendix. The strongest feature is the \wline{} line of \ws, peaking at 230 mK\@. The \wline{} spectrum for each water isotopolog is plotted in \fig{312spectra}. Thanks to the excellent signal-to-noise ratio from the long integration (8 mK rms in 0.5 km \ps{} bins), the \ws{} spectrum now shows a distinct narrow emission feature on top of the medium and broad features detected by \citet{kristensen10b}. The three components have full widths at half maximum (FWHMs) of $42.5\pm0.8$, $10.7\pm0.4$, and $3.6\pm0.3$ km \ps. \tb{linelist} lists the integrated intensities. The broad and medium FWHMs are the same as those reported by \citeauthor{kristensen10b} for lower-excitation water lines. These components arise in shocked gas and are not considered here any further.

The \we{} spectrum shows a weak emission feature ($7\sigma$) in both polarization filters, with the same intensity in the WBS and HRS recordings. The JPL and CDMS spectral line databases \citep{pickett98a,muller01a} show no other plausible features at this position in either the upper or the lower sideband. \fig{312spectra} shows a Gaussian fit with an FWHM fixed at 3.6 km \ps{} (taken from the \ws{} decomposition), because the feature is too weak for a reliable independent measurement of its width. The constrained fit has a peak intensity of $20\pm4$ mK and an integrated intensity of $78\pm11$ \mkkms{} (\tb{linelist}).

The \wsev{} spectrum in \fig{312spectra} shows an apparent emission feature blueshifted by 2 km \ps{} from the source velocity, along with four features of similar strength at larger offsets. Three of them could represent lines of CH$_2$DOH ($\eup/k=229$--294 K), which has been detected before in single-dish spectra of IRAS2A (\citealt{parise06a}; see also Figures \ref{fig:fullspecone} and \ref{fig:fullspec}). The JPL and CDMS databases offer no plausible identification for the fourth feature. We consider the \wline{} line undetected in \wsev, but our analysis in \sect{disc} is unaffected by the question of weather it is real or not. The potential CH$_2$DOH lines are not considered any further.

The only other confirmed feature in the entire spectral setting is the \ceo{} 10--9 line at 1097.163 GHz. It appears to be a blend of a narrow and a medium component, but the detection is too weak for an accurate decomposition. The intensity integrated over the full profile is $300\pm30$ \mkkms, split roughly halfway between the two components. Narrow and medium features have been detected previously in various low- and high-$J$ CO isotopolog lines \citep{yildiz10a}. The \ceo{} 10--9 line is analyzed as part of a larger sample of lines and sources by \citet{sanjose13a} and \citet{yildiz13a}.

The full spectrum in Figures \ref{fig:fullspecone} and \ref{fig:fullspec} appears to contain a few more lines, but their nature is uncertain and not of interest to this paper. We merely mention two possible identifications: \fmh{} at 1094.590 GHz ($\eup/k=526$ K) and \meoh{} at 1095.063 GHz ($\eup/k=498$ K). Both species have been detected previously in single-dish and interferometric studies of IRAS2A \citep{maret05a,parise06a,jorgensen07a}.


\subsection{Complementary Data}
\label{sec:compdata}
IRAS2A has been targeted in various other water lines with Herschel. \citet{visser12a} detected about a dozen spectrally unresolved lines with the PACS instrument (60--180 micron; $\eup/k$ up to 1750 K), attributed to shocked gas on 100-1000 AU scales and unrelated to the narrow emission analyzed here. No narrow emission has been detected in any of the water lines previously observed with HIFI \citep{kristensen10b}. Upper limits on such emission may be useful and are reported in \tb{linelist}. Another HIFI upper limit is available for \ws{} $3_{31}$--$4_{04}$ at 1894 GHz, observed alongside C$^+$ $^2$P$_{3/2}$--$^2$P$_{1/2}$ (Benz et al.\ in prep.). In most cases, the upper limit in \tb{linelist} is the $3\sigma$ limit computed from the rms noise in 0.5 km \ps{} bins. This method does not work for the \ws{} $2_{02}$--$1_{11}$ and $2_{11}$--$2_{02}$ lines, where the broad and medium emission features can hide a narrow feature stronger than three times the noise. Instead, we forced a three-Gaussian fit on these spectra, with the position and width of the narrow feature fixed at the values found for the \wline{} line. The adopted upper limit is twice the integrated intensity of the force-fitted narrow component. In doing so, we assume the broad and medium components do not shield any narrow emission. The $1_{10}$--$1_{01}$, $1_{11}$--$0_{00}$, and $2_{12}$--$1_{01}$ lines show narrow absorption due to cold water vapor in the outer envelope \citep{kristensen10b} and are excluded from our analysis.

Ground-based observations exist for two \we{} lines: \pdbi{} at 203 GHz ($\eup/k=204$ K) and $4_{14}$--$3_{21}$ at 391 GHz ($\eup/k=322$ K). The latter was undetected with APEX in a $16''$ beam, with a $3\sigma$ upper limit of 410 \mkkms{} (\tb{linelist}; F.\ Wyrowski priv.\ comm.). \citet{persson12a} detected spatially resolved compact and extended emission in the \pdbi{} line with the Plateau de Bure Interferometer (PdBI). The extended component is associated with the outflow and is ignored here. The compact component has an FWHM of $4.0\pm0.1$ km \ps, comparable to the narrow component in the \ws{} \wline{} spectrum in \fig{312spectra}. The compact PdBI component has an integrated intensity of 0.98 Jy km \ps{} or 42 \kkms. Its spatial extent was marginally resolved to a diameter of $0\farcs83$, probing material out to a radius of 100 AU\@. \citeauthor{persson12a}\ attributed the compact emission to a flattened inner envelope or pseudo-disk dominated by infall rather than rotation. 


\section{Analysis}
\label{sec:disc}


\subsection{Spherical Model}
\label{sec:sphmod}
The HIFI \wline{} lines have upper-level energies of 249 K\@. The bulk of the water in IRAS2A is present at temperatures below 100 K, where it generally exists as ice rather than vapor. The low gas-phase abundance in the cold envelope \citep[$\sim$\ten{-8};][]{kristensen10b,liu11a}, coupled with the low excitation temperatures, falls orders of magnitude shy of reproducing the observed narrow intensities \citep{vankempen08a}.

That leaves two options for the narrow \wline{} emission: a maser or the hot core. \citet{furuya03a} observed maser emission towards IRAS2A at 22 GHz, with intensities varying by an order of magnitude on timescales of about a year. The maser lines were seen sometimes at the source velocity and sometimes blue-shifted by a few km \ps, with FWHMs from 1.2 to 2.3 km \ps. The HIFI \wline{} and PdBI \pdbi{} spectra show no such blue-shifted emission and the emission at the source velocity is 2--3 times as broad as the 22 GHz maser lines. Furthermore, our two epochs of HIFI data \citep[\fig{312spectra} and][]{kristensen10b} show no signs of variability on a one-year baseline. Hence, we conclude that the narrow \wline{} emission is not associated with maser activity. \citet{persson12a} employed similar arguments to reach the same conclusion for their PdBI data.

Instead, the narrow HIFI and PdBI lines are thermal and likely originate in the inner 100 AU of IRAS2A, in quiescent gas above 100 K\@. If the circumstellar material in IRAS2A is approximated as a spherical envelope with a power-law density profile, the 100 K radius lies at 94 AU \citep{kristensen12a}. While this matches the size of the emitting region seen with the PdBI, continuum interferometry of IRAS2A has shown that the assumptions of spherical symmetry and a single power-law density profile break down inside a radius of $1\farcs5$ or 350 AU \citep{jorgensen04a,jorgensen05b}.

A simple test reveals that the spherical model also fails to reproduce the HIFI and PdBI data simultaneously. The test consists of synthesizing the \wline{} and \pdbi{} spectra from a step abundance profile, as described in detail by \citet{vankempen08a}. \citet{kristensen12a} constructed a source model for IRAS2A to fit the spectral energy distribution and sub-millimeter brightness profiles. The model envelope has a total mass of 5.1 $M_\odot$, distributed along an $r^{-1.7}$ power law. The temperatures of the gas and dust are coupled and decrease from 250 K at the inner edge (36 AU) to 10 K at the outer edge (17\,000 AU). The step abundance profile consists of a high abundance $X\q{in}$ above 100 K (at 94 AU) and a lower abundance $X\q{out}$ below 100 K\@. The HIFI and PdBI emission originate above 100 K, so the question we want to answer is whether the spherical envelope model can reproduce both data sets with a single $X\q{in}$.

The molecular excitation and line emission are computed with the 1D radiative transfer code RATRAN \citep{hogerheijde00a}. This is a standard approach for interpreting molecular line spectra; see e.g.\ \citet{yildiz10a}, \citet{liu11a}, and \citet{coutens12a} for recent examples of \ceo, HDO, and \w\@. We set the gas-to-dust ratio to 100 and use OH5 opacities \citep{ossenkopf94a}, appropriate for dust grains with thin ice mantles. The ortho/para ratio of \mh{} is thermalized and that of water is fixed at 3. Collision rates are taken from \citet{dubernet06a,dubernet09a} and \citet{daniel10a,daniel11a} as compiled in the LAMDA database \citep{schoier05a}. The outer \we{} abundance does not affect the \wline{} and \pdbi{} lines and is fixed at \scit{2}{-11}, derived from absorption seen with HIFI in lower rotational lines of water \citep{kristensen10b,liu11a}. The inner \we{} abundance is varied from \scit{3}{-10} to \scit{3}{-7} (\scit{2}{-7} to \scit{2}{-4} for \ws) to cover the full range from a ``dry'' to a ``wet'' hot core. Finally, the synthetic spectra are convolved to the appropriate beam size and compared to the available observations.

\begin{figure}[t!]
\epsscale{1.17}
\plotone{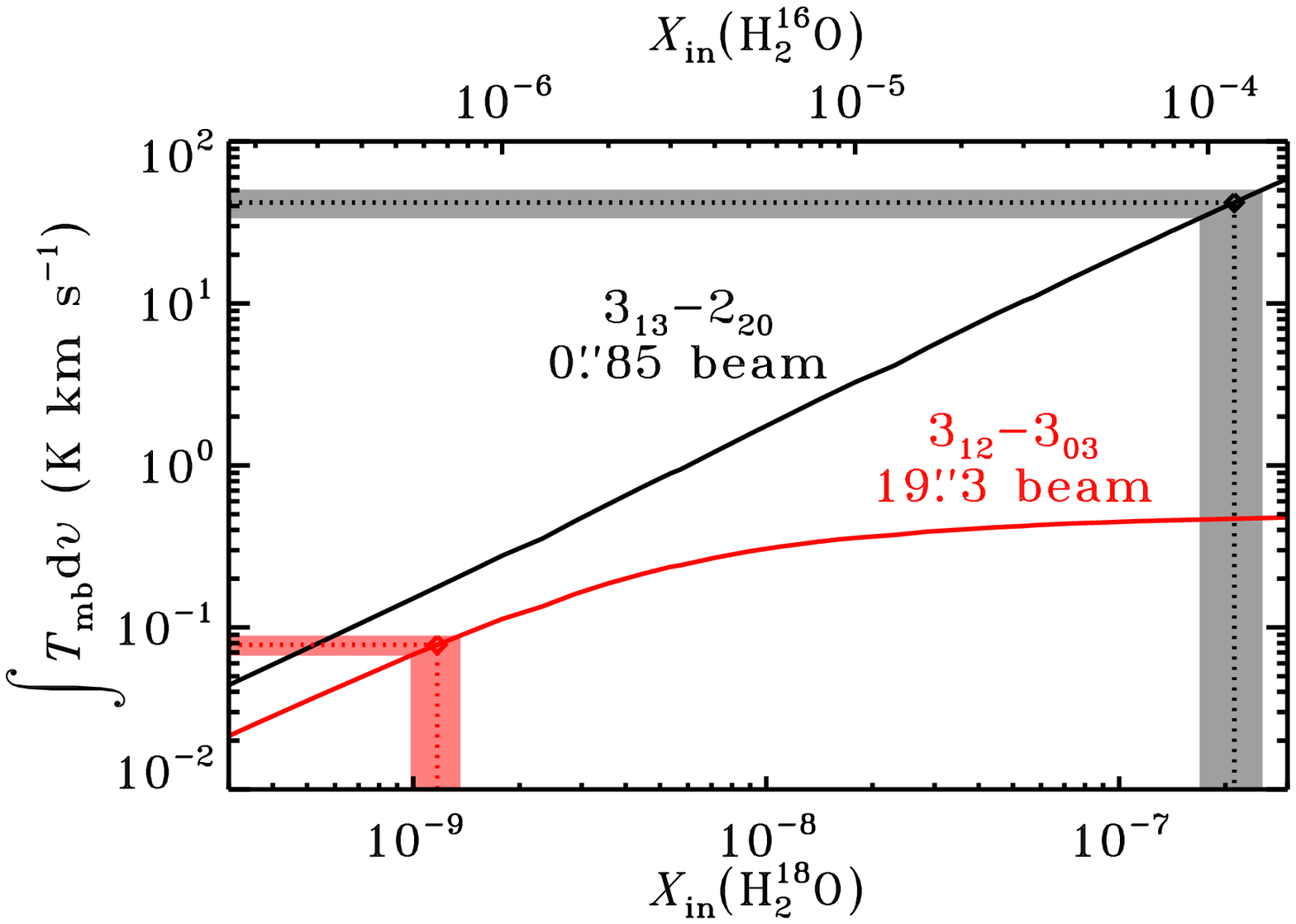}
\caption{Integrated intensities for the HIFI \wline{} line (red) and the PdBI \pdbi{} line (black) of \we{} in the spherical envelope model of \citet{kristensen12a}. The intensities, convolved to the appropriate beam size, are plotted as function of the hot core abundance of \we{} (bottom axis) or \ws{} (top axis) for a water o/p ratio of 3. The horizontal dotted lines and bars mark the observed intensities and $1\sigma$ uncertainties. The abundances required to fit each observation individually are indicated by the vertical dotted lines and bars.\label{fig:ratran}}
\end{figure}

\fig{ratran} shows the convolved intensities for the \we{} \wline{} and \pdbi{} lines as function of the inner abundance. The horizontal dotted lines and bars mark the observed intensities and $1\sigma$ uncertainties: red for \wline{} and black for \pdbi. The abundances required to fit each line individually are marked by the vertical dotted lines and bars. Within the confines of the spherical envelope model, the \we{} \wline{} spectrum from \fig{312spectra} requires an inner \we{} abundance of \scit{1.2}{-9}. However, this underproduces the observed PdBI line intensity by a factor of 200. Matching the PdBI line requires a much higher inner \we{} abundance of \scit{2.1}{-7}, but then the \wline{} line comes out a factor of 6 too strong. The estimated uncertainty on the \wline{} intensity for \we{} is 17\% (10\% calibration, 14\% statistical; Sections \ref{sec:obs} and \ref{sec:newdata}), so a factor of 6 is $35\sigma$ away. The RATRAN test clearly fails: a spherical envelope model with a single power-law density profile, such as used in previous attempts to derive hot core water abundances, cannot reproduce the single-dish and interferometric hot water observations simultaneously.


\subsection{LTE Analysis}
\label{sec:lte}
The mismatch of best-fit abundances to single-dish and interferometric data has also been seen for methanol in IRAS2A \citep{jorgensen05b} and for deuterated water in the Class 0 protostar IRAS $16293-2422$ \citep{coutens12a,persson13a}. We thus have several pieces of evidence that simple spherical models break down on the spatial scales of the hot core. Until more appropriate source models can be generated -- for example from sub-arcsecond PdBI and ALMA observations of protostellar dust and gas -- it is better to take a step back and see what we can learn about hot water in IRAS2A irrespective of the underlying density and temperature structure.


\subsubsection{PdBI Emission: Optically Thin}
\label{sec:pdbi}
In analyzing the PdBI data, \citet{persson12a} assumed optically thin emission in local thermodynamic equilibrium (LTE) and considered gas temperatures between 50 and 250 K\@. The \pdbi{} line has a critical density of a few \ten{5} \pcc. The critical density of the HIFI \wline{} line is about a factor of 1000 higher, but with densities in the inner envelope easily exceeding \ten{9} \pcc{} \citep{visser09a}, the assumption of LTE is probably justified and we adopt it here as well.

The dust and gas temperatures are likely coupled at these high densities, so temperatures below 100 K can be excluded because of freeze-out. This also means that the $0\farcs83$ emitting region resolved with the PdBI represents the entire reservoir of quiescent (i.e., non-shocked) hot gas in IRAS2A\@. The PdBI observations have a field of view of $25''$, a little larger than the HIFI beam of $19''$ at 1097 GHz. Hence, if any other hot gas had been present within the HIFI beam but outside the central $0\farcs83$, it would have been detected with the PdBI.

Following \citet{goldsmith99a}, the optical depth of a molecular line is
\begin{equation}
\label{eq:tau}
\tau_\nu = \frac{N\q{u}}{\Delta v} \frac{A\q{ul}c^3}{8\pi\nu^3} \left({\rm e}^{h\nu/kT}-1\right)\,,
\end{equation}
with $\nu$ the frequency of the transition, $A\q{ul}$ the Einstein A coefficient, and $\Delta v$ the observed FWHM\@. In LTE, the upper-level column density $N\q{u}$ is
\begin{equation}
\label{eq:nupper}
N\q{u} = \frac{N g\q{u}}{Q(T)} {\rm e}^{-\eup/kT}\,,
\end{equation}
with $N$ the total column density, $g\q{u}$ the degeneracy, and $Q$ the partition function. The PdBI line is optically thin with $\tau_\nu\approx0.1$. For a water ortho/para ratio of 3, the observed PdBI intensity requires $\cdwe\approx3\times10^{16}$ \pcs{} regardless of the exact temperature between 100 and 250 K \citep{persson12a}. The \wline{} lines have Einstein A coefficients a few thousand times faster than the \pdbi{} line (\tb{linelist}) and are optically thick for all three isotopologs. For example, at 100 K, $\tau_\nu=9.7$ for \wsev, 34 for \we, and 19\,000 for \ws\@. The optical depths decrease by about 50\% if the temperature is increased to 250 K.


\subsubsection{HIFI Emission: Optically Thick}
\label{sec:hifi}
If the HIFI emission in all three isotopologs is indeed optically thick and characterized by the same temperature, the \wline{} lines would have had the same intensities. This is clearly inconsistent with the data (\fig{312spectra}). However, the size of the optically thick region (approximately the $\tau_\nu=1$ surface) may be different for each isotopolog, leading to different beam-filling factors. As argued below, this can explain the factor of 4.4 difference in observed intensities for \ws{} and \we{} as well as the non-detection of \wsev.

\begin{figure}[t!]
\epsscale{1.17}
\plotone{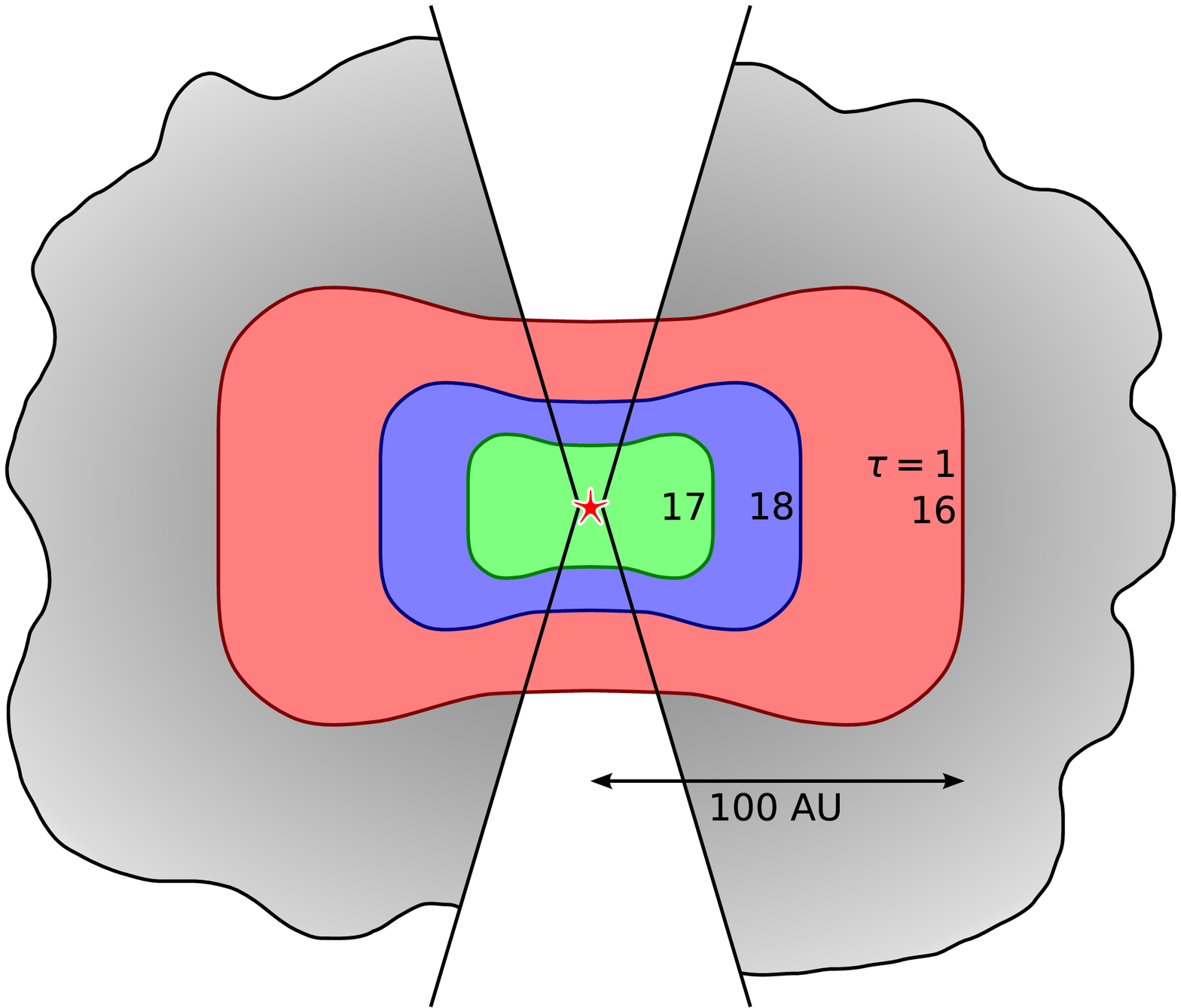}
\caption{Illustration of the hot core ($T>100$ K) in IRAS2A\@. The red and blue areas show the $\tau_\nu=1$ surfaces for the \wline{} lines of \ws{} and \we, with a difference in area that explains the different intensities observed with Herschel-HIFI (\fig{312spectra}). The green area shows the maximum size of the \wsev{} $\tau_\nu=1$ surface allowed by the $3\sigma$ upper limit from HIFI\@. The red area is the same as that of the spatially resolved \we{} \pdbi{} emission detected with the PdBI \citep{persson12a}. The hot core is embedded inside a colder envelope ($T<100$ K; shaded gray), extending out to \ten{4} AU (not drawn to scale).\label{fig:tau=1}}
\end{figure}

\fig{tau=1} shows a sketch of the flattened inner envelope of IRAS2A with the different emitting areas for the optically thick \wline{} lines. The \we{} \pdbi{} line observed with the PdBI is optically thin and traces the entire hot core to its outer radius of 100 AU\@. The extreme optical depth of 19\,000 for \ws{} \wline{} means that its $\tau_\nu=1$ surface, measured from the outside, essentially lies right at the hot core's outer edge. \ws{} therefore emits at close to 100 K\@. This is somewhat lower than the excitation temperature of 170 K assumed by \citet{persson12a}, but the difference has no substantial effect on our analysis. The temperature structure on these small scales is unknown, so we assume for simplicity that \we{} and \wsev{} also emit at 100 K\@. The radius of the \we{} $\tau_\nu=1$ surface then has to be a factor of $\sqrt{4.4}=2.1$ smaller ($r\approx47$ AU) than that of \ws{} (\fig{tau=1}). The size difference would be smaller if we take a higher temperature for \we{}, but the available data do not allow for a more precise estimate. 


\subsubsection{Water Distribution}
\label{sec:distri}
For an \we{} column density of \scit{3}{16} \pcs, the optical depth in the \we{} \wline{} line is 34 at 100 K (\eq{tau}). If the water is distributed uniformly throughout the hot core, the $\tau_\nu=1$ surface for \we{} would lie at $1/34$th of the way in from the outer edge at 100 AU, i.e., at $r\approx97$ AU\@. The surface has to be at $r\approx47$ AU to explain the observed \ws/\we{} intensity ratio from \fig{312spectra}, so the water cannot be distributed uniformly with radius.

A power-law distribution, $n({\rm H}_2{\rm O}) \propto r^{-p}$, offers a simple solution. This does not necessarily imply an abundance gradient, because it can be the total gas density itself that increases toward smaller radii \citep{visser09a}. We need to know the inner radius of the water reservoir in order to compute the power-law exponent $p$, but the inner radius cannot be measured from the PdBI data. A rough estimate would be 0.1 AU, the typical inner edge of accretion disks around T Tauri stars \citep{akeson05a}. The requirement that only $1/34$th (3\%) of the \we{} column lie at $r>47$ AU then yields $p=1.2$. The slope steepens to 1.4 if the inner radius is placed at 1 AU instead.

\begin{figure}[t!]
\epsscale{1.17}
\plotone{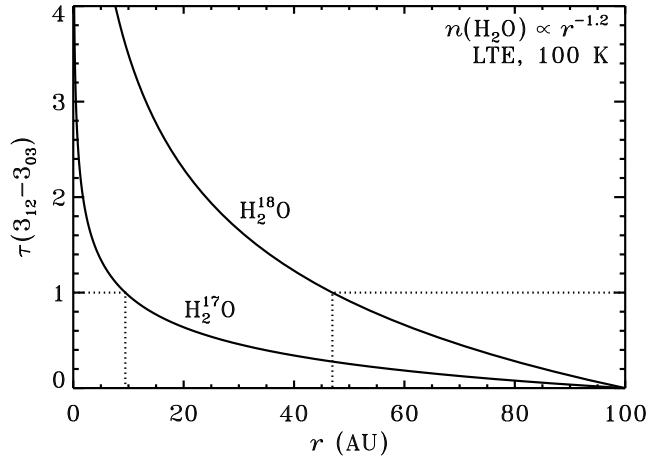}
\caption{Optical depth of \we{} and \wsev{} \wline{} measured inward from the outer edge of the hot core (\fig{tau=1}). Both curves are for power-law water density profiles at an assumed emitting temperature of 100 K\@. The dotted lines mark the $\tau_\nu=1$ radii.\label{fig:tau312}}
\end{figure}

\fig{tau312} shows the resulting optical depth profiles ($\tau_\nu$ as function of $r$) for \we{} and \wsev\@. The $\tau_\nu=1$ surface for \ws{} (not plotted) lies at 100 AU, consistent with \fig{tau=1}. \we{} reaches $\tau_\nu=1$ at 47 AU, as needed to fit the \wline{} observations. The $\tau_\nu=1$ surface for the least abundant isotopolog, \wsev, ends up at 9.4 AU, providing 110 times more beam dilution than experienced by \ws\@. This places the observable \wsev{} \wline{} intensity about a factor of 10 below the $3\sigma$ upper limit from \fig{312spectra} and \tb{linelist}.

\subsubsection{Summary of LTE Analysis}
\label{sec:columns}
In summary, the \wline{} emission observed with HIFI and the \pdbi{} emission observed with the PdBI can be explained simultaneously through an LTE analysis. The PdBI emission is spatially resolved and marks the outer boundary of the hot core in IRAS2A, with a radius of about 100 AU\@. The \pdbi{} line is optically thin and implies column densities of \scit{2}{19} \pcs{} for \ws, \scit{3}{16} \pcs{} for \we, and \scit{8}{15} \pcs{} for \wsev{} \citep{persson12a}. The 1097 GHz HIFI lines are optically thick in all three isotopologs. The observed line intensities are nonetheless different because each isotopolog becomes optically thick at a different radius, so that \ws{} experiences less beam dilution than \we{} and \wsev{} (Figures \ref{fig:tau=1} and \ref{fig:tau312}).

The precise three-dimensional distribution of water within the hot core cannot be ascertained from the HIFI and PdBI data. The simplest solution is a radial power-law dependence, $n({\rm H}_2{\rm O}) \propto r^{-p}$, with an exponent of about 1.2. We emphasize, however, that this power law is nothing more than a convenient approximation to the real 3D source structure in the hot core. The spherical radiative transfer model from \sect{sphmod} fails not because its density power law in the inner 100 AU is a little too steep, but because the assumption of spherical profiles for the density, temperature, and abundance does not hold on scales smaller than a few 100 AU. The implications of this conclusion are discussed in \sect{imp}.


\subsection{Dust Shielding and Observable Line Intensities}
\label{sec:dust}
The scenario from the previous section matches the line intensity ratios for the HIFI \wline{} isotopolog lines, but what about the absolute intensities of those and other lines? For optically thick emission in LTE, the intrinsic line intensity is the intensity of a blackbody at the gas temperature:
\begin{equation}
\label{eq:thick}
T_{\rm R}^{\rm thick} = \frac{c^2}{2 k \nu^2} B_\nu(T)\,.
\end{equation}
The optically thin equation is taken from \citet{goldsmith99a}:
\begin{equation}
\label{eq:thin}
T_{\rm R}^{\rm thin} = \frac{h c^3 N\q{u} A\q{ul}}{8 \pi k \nu^2 \Delta v} \left( \frac{1 - {\rm e}^{-\tau_\nu}}{\tau_\nu} \right)\,,
\end{equation}
with all symbols as defined as in \eq{tau}. The observable intensity $T\q{mb}$ is the intrinsic intensity $T\q{R}$ divided by the relevant beam-filling factor.

For a gas temperature of 100 K, the optically thick \ws{} \wline{} emission has an integrated intensity of 320 \kkms{} before applying beam dilution. The beam dilution factor is $(19.3/0.83)^2=540$, reducing the intensity to 590 \mkkms. This is still a factor of 1.7 stronger than observed (\tb{linelist}), which we attribute to extinction by dust. The implied optical depth of the dust is $\tau\q{d}=\ln{1.7}=0.55$ at 1097 GHz.

Single-dish and interferometric continuum data of IRAS2A provide a lower and upper bound to the observed dust optical depth. The single-dish data, processed through the spherical model of \citet{kristensen12a}, give a pencil-beam \mh{} column density of \scit{3.7}{23} \pcs{} and $\tau\q{d}=0.26$ at 1097 GHz. The interferometric data reveal an embedded pseudo-disk with a gas mass of 0.056 $M_\odot$ within $\sim$100 AU \citep{jorgensen09a}, which translates to $\tau\q{d}=2.3$. The actual optical depth encountered by molecular line emission from the hot core has to lie between these two limits, so a value of 0.55 is quite reasonable.

The final test is to check whether our hot core scenario also matches the upper limits on narrow emission in other water lines. Again assuming a gas temperature of 100 K, we repeat the above exercise to estimate intrinsic line intensities. Most lines are optically thick, so the location of their $\tau_\nu=1$ surfaces is calculated for the same $r^{-1.2}$ power-law water distribution as in \fig{tau312}. The observable line fluxes, corrected for beam dilution and dust extinction \citep[based on OH5 opacities;][]{ossenkopf94a}, are listed in \tb{linelist}. The predicted intensity of the \ws{} $2_{02}$--$1_{11}$ line at 988 GHz is right at the upper limit of how much narrow emission can be hidden in the observed broad and medium components. All other predictions fall below the upper limits. Our scenario of a 100 AU hot core with a power-law distribution of water (Figures \ref{fig:tau=1} and \ref{fig:tau312}) is therefore consistent with the combined PdBI, APEX, and Herschel data sets.

The high optical depths for the water isotopolog lines accessible with HIFI are not unique to IRAS2A\@. For example, spatially resolved data from the Atacama Large Millimeter/submillimeter Array (ALMA) show that the column density of hot water in IRAS $16293-2422$ is a factor of 30 higher than in IRAS2A \citep{persson13a}. The HIFI water lines observed in this source \citep{coutens12a} therefore likely suffer from the same problem. A combined analysis of the ALMA and HIFI data is recommended for the best overall understanding.

The conclusions from this paper can be tested with ALMA, in particular regarding the excitation and spatial distribution of hot water. The $5_{32}$--$4_{41}$ line of \we{} at 692 GHz ($\eup=728$ K) is accessible in band 9 and has been observed in the hot core of IRAS $16293-2422$ \citep{persson13a}. Within a few years, receivers in band 8 (385--500 GHz) will be able to access the $4_{14}$--$3_{21}$ and $4_{23}$--$3_{30}$ lines of \we{} at 391 and 489 GHz ($\eup/k=322$ and 430 K). All three lines are predicted to be optically thin in IRAS2A and strong enough to be observable (\tb{linelist}). ALMA's spatial resolution of $0\farcs1$ or better in bands 8 and 9 will enable an investigation of the hot water in IRAS2A at an unprecedented level of detail.


\subsection{Water Abundance and Deuteratium Fractionation in the Hot Core}
\label{sec:c18o}
\citet{persson12a} derived an \ws{} abundance of \scit{4}{-6} in the inner 100 AU of IRAS2A by comparing their \we{} column density from the PdBI to the dust mass measured in a similar beam by \citet{jorgensen09a}. However, this is not the same as the hot core abundance: the simple picture painted in \fig{tau=1} belies a more complicated three-dimensional structure, in which much of the gas and dust may be below 100 K\@. We present here an alternative approach to derive the real hot core water abundance, i.e., relative to only the gas above 100 K.

\mh{} cannot be observed directly, so CO is used as a proxy for the hot core mass. The best available tracers of the hot CO column are the optically thin 9--8 and 10--9 lines of \ceo{} ($\eup=237$ and 290 K), observed with HIFI in IRAS2A\@. There are two caveats, however: the hot CO abundance itself is uncertain by a factor of a few and some fraction of the 9--8 and 10--9 line intensities originates at temperature below 100 K \citep{yildiz10a,yildiz13a}. We address the latter issue first.

\begin{figure}[t!]
\epsscale{1.17}
\plotone{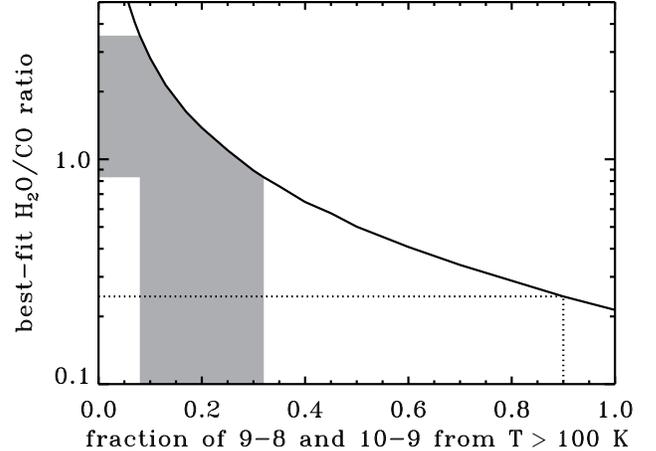}
\caption{\w/CO column density ratio in the hot core required to match the observed \ceo{} 9--8 and 10--9 line intensities if only a certain fraction of those intensities originates in gas hotter than 100 K\@. The shaded area brackets the fractions computed by \citet{yildiz10a,yildiz13a}. The dotted line marks the conservative lower limit to the \w/CO ratio discussed in the text.\label{fig:h2o/co}}
\end{figure}

The 9--8 and 10--9 lines have critical densities of about \ten{6} \pcc, so the assumption of LTE is justified. Invoking again an excitation temperature of 100 K, \eq{thin} is used to predict the \ceo{} column density required to match a certain line intensity. In the hypothetical case that all of the observed 9--8 and 10--9 emission arises in the hot core (about 150 \mkkms{} for each line; \tb{linelist}), we get $\cdceo\approx\scim{2}{17}$ \pcs{} (corrected for beam dilution and dust extinction) or an \w/CO abundance ratio of $\sim$0.2. \fig{h2o/co} shows how the derived \w/CO ratio changes if only a certain fraction of the 9--8 and 10--9 emission arises in the hot core. For example, the \w/CO ratio is 0.5 for a fraction of 50\%.

Full radiative transfer calculcations on the spherical envelope model of IRAS2A show that about 10--30\% of the observed \ceo{} 9--8 and 10--9 emission originates above 100 K \citep{yildiz10a,yildiz13a}. This range is marked by the shaded area in \fig{h2o/co} and implies an \w/CO abundance between 0.8 and 3.6. However, as noted in \sect{sphmod}, the spherical model is reliable only for radii larger than about 350 AU, where the dust temperature is $\sim$45 K \citep{kristensen12a}. This can be used for a firm limit: the real fraction of 9--8 and 10--9 emission arising above 100 K cannot be larger than the fraction arising above 45 K in the spherical radiative transfer models. That latter fraction is $\sim$65--80\% for the 9--8 line and $\sim$90\% for the 10--9 line \citep{yildiz10a,yildiz13a}. According to \fig{h2o/co}, the lower limit on the hot core \w/CO abundance ratio is then 0.25.

\citet{yildiz10a} derived a \ceo{} abundance of \scit{1.5}{-7} (\scit{8}{-5} for CO) for all gas above 25 K, a factor of 3 below the carbon and oxygen elemental abundances. This value is well constrained for the bulk of the gas between 25 and 100 K, but remains somewhat uncertain for the hot core above 100 K: the hot core represents only a small mass fraction and contributes little emission even to the \ceo{} 9--8 and 10--9 lines \citep{yildiz10a,yildiz13a}, so varying the hot core CO abundance by a factor of a few would not affect the overall line intensities significantly.

Taking the CO abundance of \scit{8}{-5} in IRAS2A at face value, the results from \fig{h2o/co} yield a lower limit to the hot core water abundance of \scit{2}{-5}. This is a conservative estimate, as the actual fraction of \ceo{} 9--8 and 10--9 emission arising above 100 K is likely to be smaller than the fraction of 90\% derived from the 45 K cutoff above. The available PdBI and HIFI observations are entirely consistent with a water abundance of (1--$2)\times10^{-4}$ in the hot core, as expected for the evaporation of water ice above 100 K \citep{rodgers03a} and as typically found in high-mass protostars \citep{vandertak06a,chavarria10a,herpin12a}. The uncertainties in the \w/CO ratio and the hot core CO abundance also allow for water abundances down to about \ten{-5}. Hot water abundances of less than \ten{-5}, as derived previously for IRAS2A \citep{kristensen10b,liu11a} or reported for other low-mass sources \citep{ceccarelli00a,maret02a,coutens12a}, are possible only if the inner CO abundance is significantly overpredicted by spherical models.

Lastly, we revisit the question of deuterium fractionation in the hot core of IRAS2A\@. \citet{liu11a} presented five rotational lines of HDO measured with single-dish facilities. The best tracer of hot HDO is the $3_{12}$--$2_{21}$ line at 226 GHz with an upper-level energy of 168 K\@. Once more invoking LTE at 100 K, the observed intensity of 0.50 \kkms{} translates to an HDO column density of \scit{2}{16} \pcs{} (\eq{thin}). \citeauthor{liu11a}\ derived a hot HDO/\w{} ratio of $\geq0.01$ based on a preliminary water abundance of $\leq10^{-5}$, which is now known to be too low. A more reliable deuterium fraction can be computed directly from the water column density of \citet{persson12a}, bypassing the uncertainties in the \w/CO ratio and the CO abundance. With $\cdws=\scim{2}{19}$, the new HDO/\w{} abundance ratio in the hot core of IRAS2A is \scit{1}{-3}. This is the same level of deuterium fractionation as in IRAS $16293-2422$ \citep{persson13a} and a factor of 2 higher than the upper limit for NGC1333 IRAS4B \citep{jorgensen10b}, both of which are based on interferometric data alone.


\subsection{Implications}
\label{sec:imp}
Continuum interferometry of IRAS2A and other low-mass embedded protostars has revealed the presence of a compact density enhancement, typically on $\sim$100 AU scales, that is not accounted for by simple spherical envelope models \citep{jorgensen04a,jorgensen05b,jorgensen07a,chiang08a}. For a given column density of hot water, correcting for the higher density would lead to a lower water abundance than what is found from the spherical models. Our conclusion in the previous section is exactly opposite: the new hot water abundance, based on the combined PdBI and HIFI data for \we{} and \ceo{}, is higher than that obtained previously from single-dish data only.

The reason for this apparent discrepancy is the temperature structure in the inner 100 AU of IRAS2A\@. With an \we{} column density of \scit{3}{16} \pcs{} and an \ws{} abundance of \scit{2}{-5} or higher, the column density of hot \mh{} is \scit{8}{23} \pcs{} or lower. The total gas column within 100 AU, based on continuum interferometry, is \scit{4}{24} \pcs{} \citep{jorgensen09a}. From this we reach the same conclusion as \citet{persson12a}: only a limited fraction of all quiescent material inside of 100 AU is hotter than 100 K\@. \citeauthor{persson12a}\ derived a hot fraction of 4\% for an assumed water abundance of \scit{1}{-4}. Given the uncertainty in the water abundance, the actual hot fraction can be anything from 2 to 20\%.

The corresponding physical picture is not that of a spherical hot core, because in such a geometry it would be difficult to have the bulk of the material at temperatures below 100 K, while still keeping some hotter gas out to 100 AU\@. An embedded disk or pseudo-disk is more likely, with a few per cent of its mass heated to above the evaporation temperature of water. This scenario is qualitatively consistent with predictions from evolutionary models \citep{brinch08a,visser09a,ilee11a,harsono13a}. Specifically, it suggests that the bulk of the disk starts cold, with little water ice evaporating as material accretes from the larger-scale envelope \citep{visser09a}.

Regarding the chemistry, the lower limit on the hot core water abundance is consistent with the value of $\sim$\ten{-4} expected from the evaporation of water ice above 100 K\@. \citet{stauber06a} presented models of hot and cold envelope chemistry in the presence of UV and X-ray fields. For an X-ray flux of \ten{-3} erg \ps{} \pcs{} (equivalent to an isotropic X-ray luminosity of \scit{3}{28} erg \ps{} at 100 AU), the hot water abundance is reduced from \ten{-4} to \ten{-6} on a timescale of \ten{4} yr. However, the gas above 100 K is essentially in freefall, leading to a dynamical timescale of perhaps only 100 yr \citep{schoier02a,visser09a}. That is still long enough for water to evaporate from the grains, but not for it to be destroyed by X-rays. Our relatively high hot water abundance can be sustained by continuous replenishment of icy grains from further out in the envelope.

Several complex organic species have been observed in IRAS2A, including \meoh, \mecn, and CH$_3$OCH$_3$ \citep{maret05a,parise06a,bottinelli07a,jorgensen07a,persson12a}. The line widths of a few km \ps{} and the excitation temperatures of about 100 K mark IRAS2A as a prototypical hot core or hot corino. This is in contrast to other protostars such as NGC1333 IRAS4A and 4B, where the observed complex organics appear to be associated with outflow-induced shocks \citep{jorgensen07a}. The aforementioned dynamical timescale of 100 yr in the hot core of IRAS2A is too short for the formation of complex organics in the gas phase after evaporation of the icy grain mantles. Instead, the organics observed in IRAS2A are likely to be so-called first-generation species \citep{herbst09a}, formed via grain-surface chemistry in the cold outer envelope and released into the gas phase in the hot core.


\section{Conclusions}
\label{sec:conc}
This paper presents Herschel-HIFI spectra of the \wline{} lines of \ws, \wsev, and \we{} (1097 GHz; $\eup/k=249$ K) in the low-mass Class 0 protostar NGC1333 IRAS2A\@. The deep 5.1-hr integration brings out a narrow emission feature (${\rm FWHM}=3.6$ km \ps) in the \ws{} spectrum on top of two broader components. The \we{} spectrum shows a narrow feature only and the line is not detected in \wsev\@. The broad components seen in \ws{} are associated with shocked gas and are not discussed here further.

Guided by spatially resolved \we{} \pdbi{} emission detected with the PdBI \citep{persson12a}, we attribute the narrow emission from the HIFI \ws{} and \we{} spectra to a hot core ($T>100$ K) with a radius of about 100 AU\@. A spherical envelope model with a power-law density profile and a step abundance for water is unable to reproduce the HIFI and PdBI intensities simultaneously. This is consistent with previous comparisons between single-dish and interferometric data \citep{jorgensen04a,jorgensen05b} and is additional proof that these simple models cannot be used for an accurate analysis of protostellar hot cores.

We present an alternative scenario where the HIFI lines are optically thick and the difference in observed intensities is due to different areas of the $\tau_\nu=1$ surfaces of the three isotopologs. The hot core is moderately shielded by dust with an optical depth of 0.55 at 1097 GHz. The beam-diluted and extinction-corrected emission from the hot core matches the intensities of the PdBI line, the two detected \wline{} HIFI lines, as well as upper limits on various other lines targeted with HIFI\@.

Because the detected HIFI emission is optically thick, the column density of water in the hot core (100 AU radius) is unchanged from the values of \scit{3}{16} \pcs{} for \we{} or \scit{2}{19} \pcs{} for \ws{} derived by \citet{persson12a}. We derive CO column densities above 100 K from \ceo{} $J=9$--8 and 10--9, and find a conservative lower limit to the \w/CO abundance ratio of 0.25. The hot CO abundance is \scit{8}{-5} \citep{yildiz10a}, so the water abundance in the hot core can be anywhere from \scit{2}{-5} to \scit{2}{-4}. Abundances of less than \ten{-5}, as previously derived from a spherical model, are only possible if the CO abundance above 100 K is significantly lower than that between 25 and 100 K\@. The revised water abundance implies a hot HDO/\w{} ratio of \scit{1}{-3}, an order of magnitude lower than earlier estimates and similar to interferometric HDO/\w{} measurements in other Class 0 protostars.


\acknowledgements
The authors thank the entire WISH team -- in particular the internal referees Asunci{\'o}n Fuente and Silvia Leurini -- for valuable discussions. Magnus Persson kindly helped with the PdBI/HIFI comparison and S{\'e}bastien Maret was quick to fix a bug in CLASS\@. We thank Friedrich Wyrowski and the APEX staff for providing the upper limit on the 391 GHz line, and Arnold Benz likewise for the 1894 GHz line.
Support for this work was provided by NASA through an award issued by JPL/Caltech.
JKJ is supported by a Lundbeck Foundation Junior Group Leader Fellowship as well as the Centre of Star and Planet Formation funded by the Danish National Reseach Foundation.
Astrochemistry in Leiden is supported by the Netherlands Research School for Astronomy (NOVA), by a Spinoza grant and grant 614.001.008 from the Netherlands Organisation for Scientific Research (NWO), and by the European Community's Seventh Framework Programme FP7/2007-2013 under grant agreement 238258 (LASSIE).
HIFI has been designed and built by a consortium of institutes and university departments from across Europe, Canada, and the United States under the leadership of SRON Netherlands Institute for Space Research, Groningen, the Netherlands and with major contributions from Germany, France, and the USA\@. Consortium members are: Canada: CSA, U.\ Waterloo; France: CESR, LAB, LERMA, IRAM; Germany: KOSMA, MPIfR, MPS; Ireland: NUI Maynooth; Italy: ASI, IFSI-INAF, Osservatorio Astrofisico di Arcetri (INAF); Netherlands: SRON, TUD; Poland: CAMK, CBK; Spain: Observatorio Astron{\'o}mico Nacional (IGN), Centro de Astrobiolog{\'\i}a (CSIC-INTA); Sweden: Chalmers University of Technology (MC2, RSS, GARD), Onsala Space Observatory, Swedish National Space Board, Stockholm University (Stockholm Observatory); Switzerland: ETH Zurich, FHNW; USA: Caltech, JPL, NHSC.


\appendix
\section{Full Spectrum}
\label{fullspec}
\begin{figure}[t!]
\epsscale{1.17}
\plotone{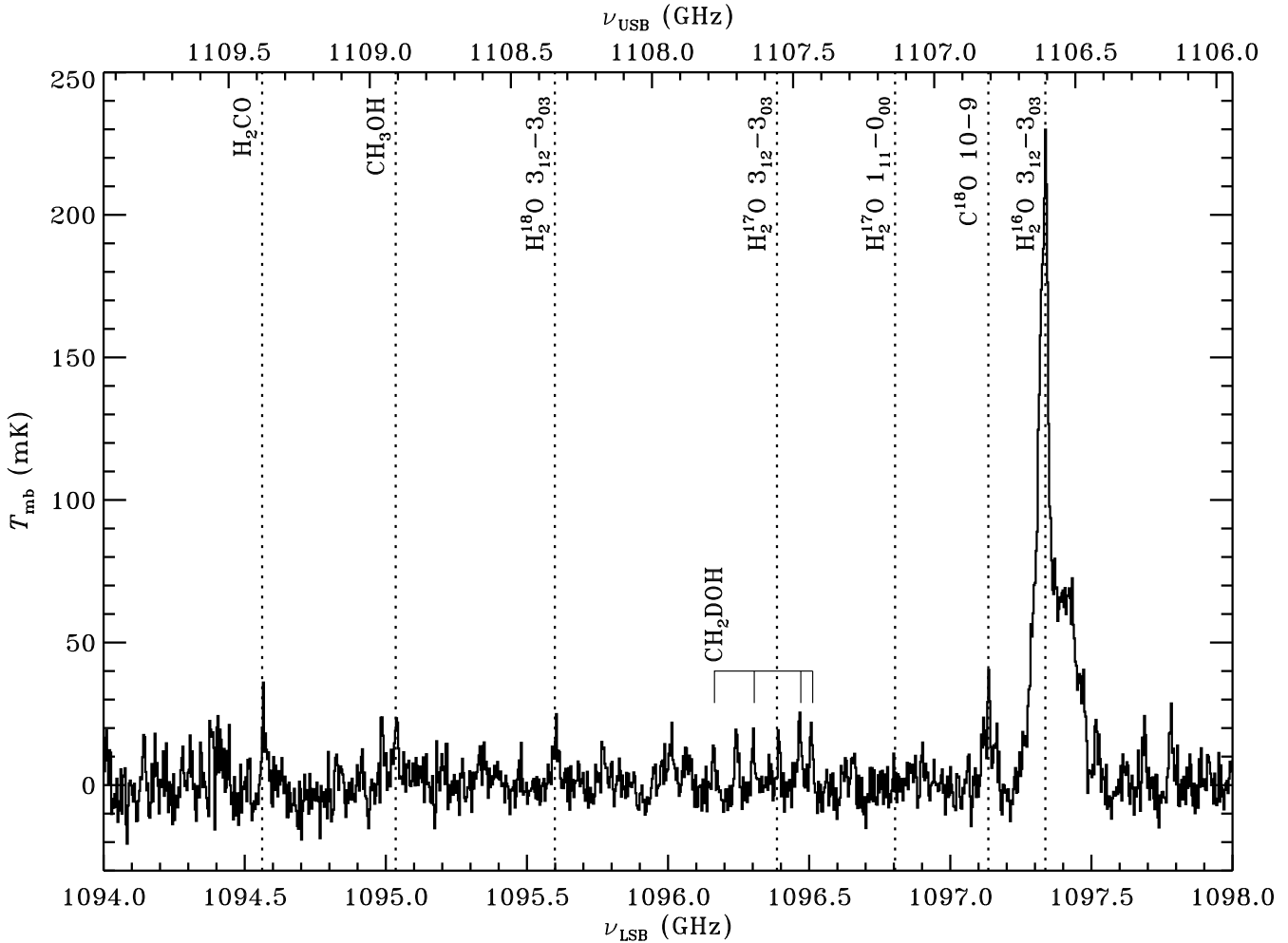}
\caption{Full baseline-subtracted HIFI spectrum for IRAS2A at 1094--1098 GHz (LSB, bottom axis) and 1106-1110 GHz (USB, upper axis) at 1.0 km \ps{} resolution. The dotted lines mark the locations of the transitions from \tb{linelist} and of some organic species, all corrected for a source velocity of 7.7 km \ps. The spectrum itself is plotted at the original observed frequencies.\label{fig:fullspecone}}
\end{figure}

\begin{figure}[t!]
\epsscale{0.89}
\plotone{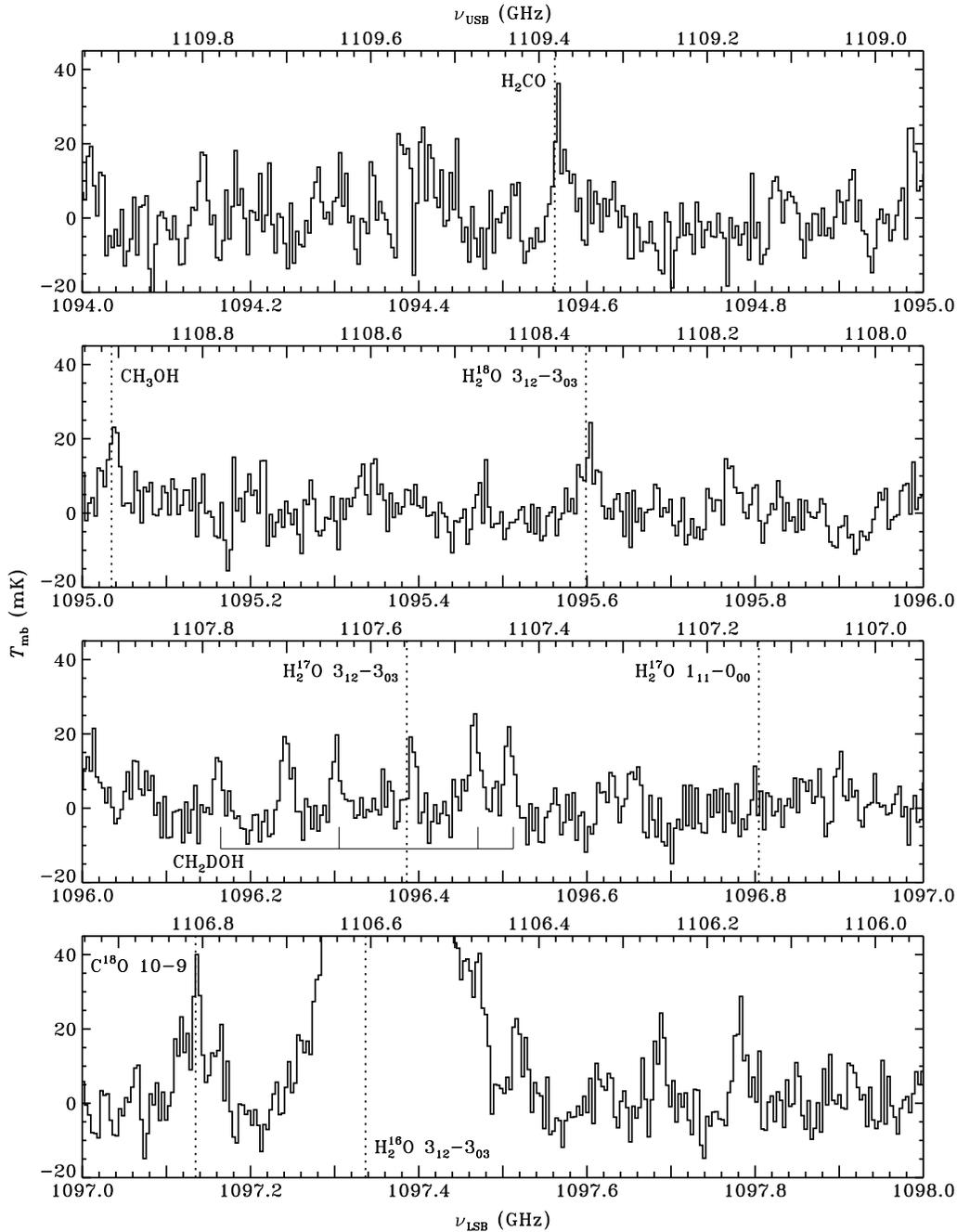}
\caption{Same as \fig{fullspecone}, but with the spectrum broken up into four chunks of 1 GHz to zoom in on the weaker features.\label{fig:fullspec}}
\end{figure}

The full spectrum from our Herschel-HIFI observations of NGC1333 IRAS2A covers frequencies from 1094 to 1098 GHz in the lower sideband (LSB) and 1106 to 1110 GHz in the upper sideband (USB). With a total integration time of 5.1 hr, this is one of the deepest spectra obtained with Herschel for a low-mass embedded protostar. The full spectrum, rebinned to a velocity resolution of 1.0 km \ps, is shown in Figures \ref{fig:fullspecone} and \ref{fig:fullspec}. The molecular line frequencies from \tb{linelist} are red-shifted by the source velocity of 7.7 km \ps{} and marked with dotted lines. Also marked are the positions of six other transitions located within the two sidebands: one of \fmh, one of \meoh, and four of CH$_2$DOH.

\ 

\end{document}